\documentstyle[aps,preprint,tighten]{revtex}
\begin{document}
\draft
\preprint{hep-th/9509005}
\title{Off-Diagonal Elements of the \\
DeWitt Expansion from the \\
Quantum Mechanical Path Integral}
\author{F.A. Dilkes
and D.G.C. McKeon\thanks{email: tmleafs@apmaths.uwo.ca
}\\
Department of Applied Mathematics \\
University of Western Ontario\\
London, Ontario, N6A 5B7\\Canada}
\date{31 August 1995) \\ (Revised 29 September 1995}
\maketitle
\begin{abstract}
The DeWitt expansion of the matrix element \linebreak[2]
\mbox{$M_{xy} = \left\langle x \right|
\exp -[\case{1}{2} (p-A)^2 + V]t \left| y \right\rangle
$},
\mbox{$(p=-i\partial)$}
in powers of $t$ can be made in a number of ways.  For $x=y$ (the
case of interest when doing one-loop calculations) numerous
approaches have been employed to determine this expansion to very
high order; when $x \neq y$ (relevant for doing calculations beyond
one-loop) there appear to be but two examples of performing the
DeWitt expansion.  In this paper we compute the off-diagonal
elements
of the DeWitt expansion coefficients using the Fock-Schwinger gauge.
Our technique is based on representing $M_{xy}$ by a quantum
mechanical
path integral.  We also generalize our method to the case of curved
space, allowing us to determine the DeWitt expansion of
$\tilde M_{xy} = \langle x|
\exp \case{1}{2} [\case{1}{\sqrt {g}}
(\partial_\mu - i A_\mu)g^{\mu\nu}{\sqrt{g}}(\partial_\nu -
i A_\nu) ] t| y \rangle
$
by use of normal coordinates.  By comparison with results for the
DeWitt expansion of this matrix element obtained by the iterative
solution of the diffusion equation, the relative merit of different
approaches to the representation of $\tilde M_{xy}$ as a
quantum mechanical path integral
can be assessed.  Furthermore, the exact dependence of
$\tilde M_{xy}$ on some geometric scalars can be determined.
In two appendices, we discuss boundary effects in the
one-dimensional
quantum mechanical path integral, and the curved space
generalization
of the Fock-Schwinger gauge.
\end{abstract}
\pacs{04.62+v, 11.10.Jj, 11.15.Bt}
\section{Introduction}
There is a long history of computing the elements $a_n(x_0,\Delta)$
in the expansion of $M_{xy}$ in powers of $t$,
\begin{mathletters}
\begin{eqnarray}
M _{xy} & \equiv & \left\langle x \right|
 \exp \left\{ - \left[ \case{1}{2}
(p-A)^2+V \right] t \right\}
\left| y \right\rangle
\label{e1a} \\
& = & \frac{e^{-\Delta^2/2t}}{(2\pi t)^{D/2}} \sum^\infty_{n=0}
a_n(x_0,\Delta) t^n \, ,  \label{e1b}
\end{eqnarray}
\end{mathletters}%
$$
(D = \text{no. of dimensions}, \; p=-i\partial, \; x_0 =
(x+y)/2, \; \Delta = x-y) \, ,
$$
in the limit $ \Delta = 0 $ \cite{r1}.  This expansion when
$\Delta = 0$
is extremely useful when examining certain properties of the
generating functional at one-loop order; in particular, the
divergence structure of a theory at one-loop order can be discerned.
Among the approaches used to evaluate $a_n(x_0,0)$ are the
perturbative solution to the heat equation \cite{r1,r2}, the use of
pseudo-differential operators \cite{r3}, working in momentum space
\cite{r4}, systematically rearranging a Schwinger expansion of
(\ref{e1a}) in powers of $A$ and $V$ into an expression of the form
(\ref{e1b}) \cite{r5,r6} and representing (\ref{e1a}) as a quantum
mechanical path integral (QMPI) hence expanding it in powers of $t$
\cite{r7,r8}.  The only places of which the authors are aware where
$a_n(x_0,\Delta)$ is considered for $\Delta \neq 0$ are in \cite{r2}
and \cite{r6}.  These coefficients are useful in considering
multi-loop
processes \cite{r9}, which motivates us to pursue them further.
The quantum mechanical path integral has proved useful in computing
Green's functions at one-loop order \cite{r10,r11,r12,r13} and
beyond
\cite{r16,r14,r15}; this suggests using this approach to examine
$a_n(x_0,\Delta)$ for $\Delta \neq 0$.  Although our method is not
identical to that of \cite{r7}, the two approaches are similar
and both results agree when $\Delta = 0$.

The representation of
$$
\tilde{M}_{xy} = \left\langle x \right|
\exp \left\{ \case{1}{2} \left[
\case{1}{\sqrt{g}} (\partial_\mu -iA_\mu) g^{\mu\nu}\sqrt{g}
(\partial_\nu -iA_\nu)
\right]t \right\} \left| y \right\rangle
$$
in terms of a quantum mechanical path integral is not uniquely
specified \cite{r17,r18,r19}, as discussed in \cite{r16}.  We
use one of the various forms of the QMPI to expand $\tilde{M}_{xy}$
and compare our results with those  of \cite{r2}.  Furthermore, a
partial summation of the DeWitt expansion to obtain the full
dependence of $\tilde M_{xy}$ on $R$ and $R_{\alpha\beta}
\Delta^\alpha \Delta^\beta$ is possible \cite{r20}.

\section{Expanding $M_{xy}$}
\label{flat}
It is possible to represent $M_{xy}$ as an appropriately normalized
 QMPI \cite{r21},
\begin{equation}
\label{e3}
M_{xy} = \int^x_y Dq(\tau) \, {\cal P} \exp \int^t_0 d\tau \left\{
-\frac{\dot{q}^2(\tau)}{2} + i \dot{q}(\tau) \cdot A(q(\tau)) -
V(q(\tau))\right\}
\end{equation}
where path-ordered integration is implied over trajectories with end
points $q(0) = y$ and $q(t) = x$.

We attempt to construct a power
series about some point $x_0$ which we arbitrarily choose to be the
mid-point between $x$ and $y$.  Defining the relative coordinate
$\delta$ by
\begin{equation}
\label{e4}
q(\tau) = x_0 + \delta(\tau) \, ,
\end{equation}
and imposing the Fock-Schwinger gauge condition \cite{r22},
\begin{equation}
\label{e5}
\delta(\tau) \cdot A(x_0 + \delta(\tau)) = 0 \, ,
\end{equation}
one can expand the gauge field in powers of $\delta$,
\begin{mathletters}%
\begin{eqnarray}
\label{e6a}
A_\mu(x_0 + \delta(\tau)) & = & \int^1_0 d\alpha \, \alpha \,
\delta^\lambda (\tau) F_{\lambda\mu}(x_0 + \alpha \delta(\tau)) \\
\label{e6b}
& = & \sum^{\infty}_{N=0} \frac{1}{N!(N+2)} \left[ \delta(\tau)
\cdot
D(x_0) \right]^N \delta^\lambda(\tau) F_{\lambda\mu}(x_0) \, .
\end{eqnarray}
\end{mathletters}%
The scalar potential can be similarly expanded,
\begin{equation}
\label{e7}
V(x_0+\delta(\tau)) = \sum^\infty_{N=0} \frac{1}{N!} \left[
\delta(\tau) \cdot D(x_0) \right] ^N V(x_0) \, .
\end{equation}
Here gauge-covariant differentiation at $x_0$ has been denoted by
$D(x_0)$.
Together, (\ref{e6b}) and (\ref{e7}) allow (\ref{e1a}) to be written
as
\begin{eqnarray}
\label{e8}
M_{xy} & = & \int^{\Delta/2}_{-\Delta/2}
D\delta(\tau) \exp \left[ - \int^t_0 d\tau
\frac{\dot\delta^2(\tau)}{2}
\right] \sum^\infty_{L=0} \frac{1}{L!} \\
& & \times {\cal P} \left\{ \sum^\infty_{N=0}
\frac{1}{N!} \int^t_0 d\tau [\delta(\tau) \cdot D(x_0)]^N \left[
\frac{i}{N+2} \dot{\delta}^\mu(\tau)
\delta^\lambda(\tau) F_{\lambda\mu}(x_0) - V(x_0) \right]
\right\} ^L \, . \nonumber
\end{eqnarray}
In the above expansion we intend to treat all
terms in the expansions of the potentials as
perturbations on the free-field action $\case{1}{2} \int^t_0
d\tau \dot\delta^2(\tau)$.
By contrast, the authors of reference~\cite{Igor}
have shown how, through
the introduction of an appropriate tensor basis, one can
include the lowest-order term of the derivative expansion of
the electromagnetic field in a non-perturbative fashion.
Although the latter
technique could be applied here,
we adopt the purely perturbative approach which
is algebraically simpler and more suitable for the purpose of
illustration; furthermore, we are free to consider a non-abelian
gauge group.

The path integral in (\ref{e8}) can be evaluated by systematic
functional
differentiation of the standard result \cite{r21,r23,r12}
\begin{eqnarray}
& & \int^{\Delta/2}_{-\Delta/2} D\delta(\tau) \exp \int^t_0 d\tau
\left\{ -\frac{\dot\delta^2(\tau)}{2} + \gamma(\tau) \cdot
\delta(\tau) \right\}  \label{e9}\\
& &
=  \frac{e^{-\Delta^2/2t}}{(2 \pi t)^{D/2}} \exp \bigg\{ \int^t_0
d\tau \left( -\frac{1}{2} + \frac{\tau}{t} \right) \Delta \cdot
\gamma(\tau) - \case{1}{2} \int^t_0 d\tau \, d\tau' G(\tau,\tau')
\gamma(\tau) \cdot \gamma(\tau') \bigg\} \, , \nonumber
\end{eqnarray}
with respect to $\gamma_\alpha(\tau)$ and then setting
$\gamma = 0$.
(Here, $ G(\tau,\tau') \equiv \case{1}{2} |\tau - \tau'| -
\case{1}{2} (\tau + \tau') + \frac{\tau\tau'}{t} $
is the Green's function of
a free particle on the worldline.\footnote{In appendix \ref{PI} we
address some concerns about the validity of this Green's function
on the finite interval $[0,t]$.
The complications presented there are not expected to contribute in
this flat-space limit.})
For example, after two such derivatives, it is easily shown that
\begin{eqnarray}
& & \int^{\Delta/2}_{-\Delta/2} D\delta(\tau) \,
\delta^\alpha(\tau_a)
\delta^\beta(\tau_b) \exp \left[ - \int^t_0 d\tau
\frac{\dot\delta^2(\tau)}{2} \right]  \label{e10} \\
& & = \frac{e^{-g_{\mu\nu}\Delta^\mu \Delta^\nu
/2t}}{(2\pi t)^{D/2}}
\left[- G(\tau_a,\tau_b) g^{\alpha\beta} +
\left( -\frac{1}{2} +
\frac{\tau_a}{t} \right) \Delta^\alpha \left( -\frac{1}{2} +
\frac{\tau_b}{t} \right) \Delta^\beta \right] \, . \nonumber
\end{eqnarray}

{}From (\ref{e9}), it is easily seen that no term in (\ref{e8})
will involve factors of $\Delta^2$.  From this observation,
combined with simple power counting
arguments, it is straightforward to tabulate the possible
contributions to the various coefficients $a_n(x_0,\Delta)$ (see
table~\ref{t1}).  (When a temperature dependent QMPI is considered
as
in \cite{r12}, then the temperature provides a second dimensionful
parameter which must be considered).
The coefficients of the various contributions can be easily
determined by
appropriately choosing $L$ and $N$ in (\ref{e8}) and
then systematically
applying (\ref{e9}).  For example, if $L=n, N=0$, then it is
apparent that the contribution to $a_n(x_0,\Delta)$ proportional to
$V^n(x_0)$ is $\frac{1}{n!} (-V(x_0))^n $.  By setting $L=N=1$ in
(\ref{e8}), we find after a very short calculation that the
contribution to $a_1(x_0,\Delta)$ proportional to $\Delta^\alpha
D_\beta {F_\alpha}^\beta$ is $\frac{-i}{12} \Delta^\alpha
D_\beta {F_\alpha}^\beta $.  With $L=2, N=0$, the contribution to
$M_{xy}$ is straightforwardly computed to be
$$
M^{(2,0)}_{xy} = \frac{e^{-\Delta^2/2t}}{(2\pi t)^{D/2}} \left(
-\frac{t}{24} \Delta^\alpha \Delta^\beta F_{\alpha\mu}
{F_\beta}^\mu
- \frac{t^2}{48} F_{\alpha\beta} F^{\alpha\beta} \right) \, ,
$$
giving a contribution to both $a_1(x_0,\Delta)$ and
$a_2(x_0,\Delta)$.  These results are all consistent with the flat
space limit of the expressions for $a_n(x_0,\Delta)$ given in
\cite{r2}.

\section{Expanding ${\tilde M_{xy}}$}

As has been noted in the introduction and in \cite{r16}, there
are various
representations of the matrix element
\begin{equation}
\label{e13}
\tilde{M}_{xy} = \left\langle x \right|
\exp \left\{ \case{1}{2} \left[
\case{1}{\sqrt{g}}(\partial_\mu -iA_\mu) g^{\mu\nu} \sqrt{g}
(\partial_\nu -iA_\nu)
\right]t \right\} \left| y \right\rangle
\end{equation}
in terms of a QMPI.  We adopt the approach of \cite{r19} in which
the QMPI is
computed using a normal coordinate expansion of the coordinate being
integrated over and the form of the classical action is fixed by
having $\tilde M_{xy}$ satisfy the appropriate heat kernel equation.
(This representation does not coincide with the expression given in
\cite{r24} for any value of the parameter $p$ appearing there.)
This representation gives a
dependence of $\tilde M_{xy}$ on $R$ that coincides with that of
\cite{r20,r2}, and agrees, to all orders so far checked,
with the results of
\cite{r2}.  The full dependence of $\tilde M_{xy}$ on
$R_{\alpha\beta} \Delta^\alpha \Delta^\beta$ can also be determined.

We are now faced with evaluating
\begin{eqnarray}
\tilde M_{xy} & = & \int^x_y Dq^\alpha(\tau) {\sqrt{g(q(\tau))}}
\, {\cal P} \exp
\bigg\{ - \int^t_0 d\tau \Big[ \case{1}{2} g_{\mu\nu}(q(\tau))
\dot q^\mu(\tau) \dot q^\nu(\tau)  \label{e14} \\
& & \hspace{3cm}
-i \dot q^\mu(\tau)
A_\mu(q(\tau)) + \case{1}{8} R(q(\tau)) \Big]
\bigg\} \, . \nonumber
\end{eqnarray}
The factor $\sqrt{g(q(\tau))}= \det^{1/2} g_{\mu\nu}(q(\tau))$
in the measure also occurs in the
non-linear sigma model \cite{r25}, but there it is usually discarded
as it gives a contribution to the effective action that is
proportional to $\delta(0)$ which, when regulated using dimensional
regularization \cite{r26} or operator regularization \cite{r32},
goes to zero.
We are dealing with a model in which no regularization is
required; indeed it turns out that divergent contributions from
$\sqrt{g}$ are
essential to render the path integral in (\ref{e14}) well-defined.
It is most convenient to incorporate the effects of $\sqrt g$ by
using ghosts as was done in \cite{r19}.  Perhaps the simplest way to
do this is to introduce a vector of real Bosonic fields
$b_\alpha(\tau)$ which vanish at the end points ($\tau = 0,t$).
(The structure of the ghost sector is not unique, and the authors of
\cite{r19} opt, instead, to use either a pair of Fermionic scalars
or a contravariant Bosonic ghost $b^\alpha$ and two
contravariant Fermionic ghosts $c^\alpha$ and $\overline c^\alpha$.)
Then, equation (\ref{e14}) can be re-expressed as
\begin{eqnarray}
\label{e15}
\tilde M_{xy} & = & \int^x_y Dq^\alpha(\tau) \int^0_0
Db_\beta(\tau) {\cal P} \exp
\bigg\{ - \int^t_0 d\tau \Big[
\case{1}{2} g^{\mu\nu} b_\mu(\tau) b_\nu(\tau)
\\
& & \hspace{2cm}
+ \case{1}{2} g_{\mu\nu}(q(\tau)) \dot q^\mu(\tau) \dot q^\nu(\tau)
-i \dot q^\alpha(\tau)
A_\alpha(q(\tau)) + \case{1}{8} R(q(\tau)) \Big]
\bigg\} \, . \nonumber
\end{eqnarray}

A normal coordinate expansion \cite{r27} is now made about a point
$\phi(\tau)$ so that
\begin{equation}
\label{e16}
q^\alpha(\tau) = \phi^\alpha(\tau) + \pi^\alpha(\xi(\tau)).
\end{equation}
Following the developments of \cite{r26}, we have
\begin{mathletters}
\begin{equation}
\label{e17a}
R(q(\tau)) = R(\phi(\tau)) + \case{1}{1!} R_{;\alpha}(\phi(\tau))
\xi^\alpha(\tau)
+ \case{1}{2!} R_{;\alpha\beta}(\phi(\tau)) \xi^\alpha(\tau)
\xi^\beta(\tau)
+ \cdots \, ,
\end{equation}
\begin{eqnarray}
g_{\mu\nu}(q(\tau)) & = & g_{\mu\nu} (\phi(\tau)) + \case{1}{3}
R_{\mu\alpha\beta\nu} (\phi(\tau)) \xi^\alpha(\tau) \xi^\beta(\tau)
+ \case{1}{6} R_{\mu\alpha\beta\nu;\gamma} (\phi(\tau))
\xi^\alpha(\tau)
\xi^\beta (\tau) \xi^\gamma(\tau)  \label{e17b} \\
& & + \left( \case{1}{20}
R_{\mu\alpha\beta\nu;\gamma\delta}(\phi(\tau)) +
\case{2}{45}R_{\mu\alpha\beta\sigma}
(\phi(\tau)){R^\sigma}_{\gamma\delta\nu}
(\phi(\tau)) \right) \xi^\alpha(\tau) \xi^\beta (\tau)
\xi^\gamma(\tau) \xi^\delta(\tau)\nonumber \\
& & + \cdots \nonumber \, ,
\end{eqnarray}
\begin{eqnarray}
g^{\mu\nu}(q(\tau)) & = & g^{\mu\nu} (\phi(\tau)) - \case{1}{3}
{{R^\mu}_{\alpha\beta}}^\nu (\phi(\tau)) \xi^\alpha(\tau)
\xi^\beta(\tau) - \case{1}{6} {{{R^\mu}_{\alpha\beta}}^\nu}_{;\gamma}
(\phi(\tau)) \xi^\alpha(\tau)
\xi^\beta (\tau) \xi^\gamma(\tau) \\
& & + \left( - \case{1}{20}
{{{R^\mu}_{\alpha\beta}}^\nu}_{;\gamma\delta}(\phi(\tau)) +
\case{3}{45}{R^\mu}_{\alpha\beta\sigma}
(\phi(\tau)){{{R^\sigma}_{\gamma\delta}}^\nu}
(\phi(\tau)) \right) \xi^\alpha(\tau) \xi^\beta (\tau)
\xi^\gamma(\tau) \xi^\delta(\tau) \nonumber \\
& & + \cdots \nonumber \, ,
\end{eqnarray}
and
\begin{eqnarray}
\dot q^\mu(\tau) & = & \dot \phi^\mu (\tau) + D_\tau
\xi^\mu(\tau) + \case{1}{3} {R^\mu}_{\alpha\beta\gamma} (\phi(\tau))
\xi^\alpha(\tau) \xi^\beta(\tau) \dot \phi^\gamma(\tau) + \cdots \,
, \label{e17c} \\
& & \left( D_\tau \xi^\mu(\tau) \equiv \dot \xi^\mu(\tau) +
\Gamma^\mu_{\beta\gamma} \xi^\beta(\tau) \dot \phi^\gamma(\tau)
\right) \, . \nonumber
\end{eqnarray}
As shown in appendix~\ref{gauge}, by imposing a gauge condition
analogous to (\ref{e5}), one finds that
the corresponding normal coordinate expansion for the gauge field is
(\ref{AexpansionF})
\begin{eqnarray}
\label{e17d}
A_\mu(q(\tau)) & = &
\case{1}{2} F_{\alpha\mu}(\phi(\tau)) \xi^\alpha(\tau)
+ \case{1}{3} F_{{\alpha}\mu;\beta}(\phi(\tau))
\xi^{\alpha}(\tau) \xi^{\beta}(\tau) \\
& &
+ \left( \case{1}{8} F_{{\alpha}\mu;\beta\gamma}(\phi(\tau))
+ \case{1}{24}
F_{{\alpha}\sigma}(\phi(\tau))
{R^\sigma}_{\beta \gamma \mu}(\phi(\tau))
\right) \xi^{\alpha}(\tau) \xi^{\beta}(\tau) \xi^{\gamma}(\tau)
\nonumber \\
& & + \left( \case{1}{30}
F_{{\alpha}\mu;\beta\gamma\delta} (\phi(\tau))
+ \case{1}{60} F_{{\alpha}\sigma}(\phi(\tau))
{R^\sigma}_{\beta\gamma\mu;\delta} (\phi(\tau))
\right.
\nonumber \\
& &
\hspace{2cm} \left. + \case{1}{30}
F_{{\alpha}\sigma;\delta}(\phi(\tau))
{R^\sigma}_{\beta \gamma \mu}(\phi(\tau))
\right)
\xi^{\alpha}(\tau) \xi^{\beta}(\tau) \xi^{\gamma}(\tau)
\xi^{\delta}(\tau) + \cdots \, .\nonumber
\end{eqnarray}
\end{mathletters}%
If we take $\phi(\tau)$ to be the geodesic mid-point, $x_0$, between
$x$ and $y$, then $\dot \phi^\alpha(\tau)$ vanishes and the above
expansions simplify a bit.  Letting $\Delta$ denote the difference
between the normal coordinates of $x$ and $y$, equation (\ref{e15})
becomes
\begin{eqnarray}
\label{e18}
\tilde M_{xy} & = & \int^{\Delta/2}_{-\Delta/2} D\xi^\alpha(\tau)
\int^0_0 Db_\beta(\tau)
\exp \left\{ - \int^t_0 d\tau \left[
\case{1}{2} g_{\mu\nu} \dot\xi^\mu(\tau) \dot\xi^\nu(\tau) +
\case{1}{2} g^{\mu\nu} b_\mu (\tau) b_\nu(\tau) \right]
\right \} e^{-Rt/8}
\\
& & \times {\cal P}
\sum^\infty_{N=0} \frac{1}{N!}
\bigg\{ - \int^t_0 d\tau \Big[ \case{1}{2} \left( \case{1}{3}
R_{\mu\alpha\beta\nu}  \xi^\alpha(\tau) \xi^\beta(\tau)
+ \case{1}{6} R_{\mu\alpha\beta\nu;\gamma}
\xi^\alpha(\tau)
\xi^\beta (\tau) \xi^\gamma(\tau) + \cdots
\right) \dot\xi^\mu(\tau) \dot\xi^\nu(\tau)
\nonumber \\
& & \hspace{3.5cm} + \case{1}{2} \left(
- \case{1}{3}
{{R^\mu}_{\alpha\beta}}^\nu \xi^\alpha(\tau)
\xi^\beta(\tau) - \case{1}{6}
{{{R^\mu}_{\alpha\beta}}^\nu}_{;\gamma}
\xi^\alpha(\tau)
\xi^\beta (\tau) \xi^\gamma(\tau) + \cdots \right) b_\mu(\tau)
b_\nu(\tau) \nonumber \\
& & \hspace{3.5cm} -i \left(
\case{1}{2} F_{\alpha\mu} \xi^\alpha(\tau)
+ \case{1}{3} F_{{\alpha}\mu;\beta}
\xi^{\alpha}(\tau) \xi^{\beta}(\tau) +
\cdots \right) \dot\xi^\mu \nonumber \\
& & \hspace{3.5cm}
+ \case{1}{8} \left( R_{;\alpha} \xi^\alpha(\tau)
+ \case{1}{2} R_{;\alpha\beta} \xi^\alpha(\tau) \xi^\beta(\tau)
+ \cdots\right) \Big] \bigg\}^N
\, . \nonumber
\end{eqnarray}
(All geometrical and gauge quantities in (\ref{e18}) are evaluated
at $x_0$.)

The standard results
\begin{mathletters}%
\begin{eqnarray}
& & \int^{\Delta/2}_{-\Delta/2} D\xi^\alpha(\tau) \exp \left\{
\int^t_0 d\tau \left( - \case{1}{2} g_{\mu\nu} \dot\xi^\mu(\tau)
\dot\xi^\nu(\tau) +
\xi^\mu(\tau) \gamma_\mu(\tau) \right) \right\}
\label{e19a} \\
& = & \frac{e^{-g_{\mu\nu}\Delta^\mu \Delta^\nu
/2t}}{(2\pi t)^{D/2}
\sqrt g} \exp \left\{ \int^t_0 d\tau \left( -
\frac{1}{2} + \frac{\tau}{t} \right) \Delta^\mu
\gamma_\mu(\tau) -
\case{1}{2} \int^\tau_0 d\tau d\tau' G(\tau, \tau') g^{\mu\nu}
\gamma_\mu(\tau) \gamma_\nu(\tau') \right\} \, , \nonumber
\end{eqnarray}
and,
\begin{eqnarray}
\label{e19b}
& & \int^0_0 Db_\alpha(\tau) \exp \left\{ \int^t_0
d\tau \left( - \case{1}{2} g^{\mu\nu} b_\mu(\tau) b_\nu(\tau) +
b_\mu(\tau) B^\mu(\tau) \right) \right\} \\
& & \hspace{1cm}
= {\sqrt g} \exp \left\{ \case{1}{2} \int^t_0 d\tau d\tau'
G^{\text{ghost}}(\tau,\tau') g_{\mu\nu}
B^\mu(\tau) B^\nu(\tau') \right\}_{B(0) = B(t) = 0} \, , \nonumber
\end{eqnarray}
\end{mathletters}%
permit one to compute the functional integrals
appearing in (\ref{e18}).  Equivalently, one can generate the
necessary contractions using the formalism of
equation~(\ref{Delta0}).  In either case, one should be careful
to use the Green's functions of
equations~(\ref{FieldProp}) (or in simple cases,
equations~(\ref{summedProp})) as discussed in appendix~\ref{PI}.

For example, if we restrict our attention to the
contribution to (\ref{e18}) that is linear in
$R_{\mu\alpha\nu\beta}$, we have
\begin{eqnarray}
\tilde M^R_{xy} & = & \int^{\Delta/2}_{-\Delta/2} D\xi^\alpha(\tau)
\int^0_0 Db_\beta(\tau) e^{-Rt/8} \nonumber \\
& & \hspace{1cm} \times \exp \left\{ - \int^t_0 d\tau
\left[ \case{1}{2} g_{\mu\nu} \dot\xi^\mu(\tau) \dot\xi^\nu(\tau)
+ \case{1}{2} g^{\mu\nu} b_\mu(\tau) b_\nu(\tau) \right]
\right\} \nonumber \\
& & \hspace{1cm} \times \left\{
- \case{1}{6} R_{\mu\alpha\beta\nu}
\int^t_0 d\tau \xi^\alpha(\tau) \xi^\beta(\tau) \dot\xi^\mu(\tau)
\dot\xi^\nu(\tau)
+\case{1}{6} {{R^\mu}_{\alpha\beta}}^\nu \int^t_0 d\tau
\xi^\alpha(\tau)\xi^\beta(\tau) b_\mu(\tau) b_\nu(\tau) \right\}
\nonumber \\
& = & \frac{e^{-g_{\mu\nu}\Delta^\mu \Delta^\nu
/2t}}{(2\pi t)^{D/2}}
e^{-Rt/8} \int^t_0
d\tau \bigg\{ -\case{1}{2} \Big[ g^{\mu\nu} g^{\alpha\beta}
G(\dot\tau,\dot\tau) G(\tau,\tau) + (g^{\mu\alpha} g^{\nu\beta} +
g^{\mu\beta}g^{\nu\alpha} ) G^2(\tau,\dot\tau)
\nonumber \\
& & \hspace{1cm}
- \Delta^\mu \Delta^\nu {\left( \case{1}{t} \right) }^2
g^{\alpha\beta} G(\tau,\tau)
- \Delta^\alpha \Delta^\beta g^{\mu\nu} \left(
- \case{1}{2} + \case{\tau}{t} \right)^2 G(\dot\tau,\dot\tau)
\nonumber \\
& & \hspace{1cm}
 - \left( \Delta^\mu \Delta^\alpha g^{\nu\beta} + \Delta^\nu
\Delta^\beta g^{\mu\alpha} + \Delta^\mu \Delta^\beta g^{\nu\alpha}
+ \Delta^\nu \Delta^\alpha g^{\mu\beta} \right) \case{1}{t}
\left( -\case{1}{2} + \case{\tau}{t} \right) G(\tau,\dot\tau)
\nonumber \\
& & \hspace{1cm}
+ \Delta^\alpha \Delta^\beta \Delta^\mu \Delta^\nu \left(
- \case{1}{2} + \case{\tau}{t} \right) ^2 \left( \case{1}{t} \right)
^2 \Big] \nonumber \\
& & \hspace{0.5cm}
+ \left[ -G(\tau,\tau) g^{\alpha\beta} + \left( -\case{1}{2} +
\case{\tau}{t} \right)^2 \Delta^\alpha
\Delta^\beta \right] G^{\text{ghost}}(\tau,\tau) g^{\mu\nu} \bigg\}
\left( \frac{1}{6} R_{\mu\alpha\beta\nu} \right)
\, .\label{e20}
\end{eqnarray}
Here, a dot over an argument of the Green's function $G$ indicates
differentiation with respect to that argument.  As explained in
appendix \ref{PI}, these derivatives cannot always be obtained at
the end points ($\tau=0,t$) by
naive differentiation of $G(\tau,\tau')$ (c.f. equation
(\ref{Gdotdot})).  When the explicit form of $G(\tau,\tau')$ and
its derivatives are substituted into
(\ref{e20}), we are left with with
\begin{eqnarray}
\tilde M^R_{xy} & = & \frac{e^{-g_{\mu\nu}\Delta^\mu \Delta^\nu
/2t}}{(2\pi t)^{D/2}}
\frac{1}{6}
e^{-Rt/8} \int^t_0 d\tau
\nonumber \\
& & \hspace{0.5cm}
\times \bigg\{
R_{\alpha\beta} \Delta^\alpha \Delta^\beta
\left[
\case{1}{t^2} \left( -\tau + \case{\tau^2}{t} \right)
- \case{1}{t} \left(
- \case{1}{2} + \case{\tau}{t} \right) ^2
- \left[ \delta(2\tau) +
\delta(2t-2\tau) \right]
\left( - \case{1}{2} + \case{\tau}{t} \right) ^2
\right]
\nonumber \\
& & \hspace{1.3cm} +
R \left[ - \case{1}{t} \left( -\tau + \case{\tau^2}{t}
\right) +
\left( -\case{1}{2} + \case{\tau}{t} \right) ^2
+ \left[\delta (2\tau) + \delta (2t-2\tau) \right]
\left( -\tau + \case{\tau^2}{t} \right)
\right]
\bigg\}
\nonumber
\\
& = & \frac{e^{-g_{\mu\nu}\Delta^\mu \Delta^\nu
/2t}}{(2\pi t)^{D/2}}
e^{-Rt/8} \left[ - \case{1}{12} R_{\alpha\beta}
\Delta^\alpha \Delta^\beta + \case{1}{24} Rt \right] \, ,
\label{e21}
\end{eqnarray}
where $R_{\alpha\beta} = {R^\mu}_{\alpha\beta\mu}$ and $R =
R_{\alpha\beta} g^{\alpha\beta}$.
(Notice that all dependence on $\delta(0)$ in (\ref{e20}) has
cancelled out due to the
compensating contributions from the ghost-fields.)
It is easily seen that the entire dependence of $\tilde M_{xy}$ on
$R$ and $R_{\alpha\beta} \Delta^\alpha \Delta^\beta$ is given by
\begin{eqnarray}
\tilde M^{(R,\Delta \cdot R \cdot \Delta)}_{xy} & = &
\frac{e^{-g_{\mu\nu}\Delta^\mu \Delta^\nu
/2t}}{(2\pi t)^{D/2}}
\sum^\infty_{N=0} \frac{e^{-Rt/8}}{N!} \left[ \frac{-R_{\alpha\beta}
\Delta^\alpha \Delta^\beta}{12} + \frac{Rt}{24} \right] ^N
\nonumber \\
& = & \frac{e^{-g_{\mu\nu}\Delta^\mu \Delta^\nu
/2t}}{(2\pi t)^{D/2}}
\exp \left[ -\frac{Rt}{12} -
\frac{R_{\alpha\beta} \Delta^\alpha \Delta^\beta}{12} \right]
\label{e22} \, .
\end{eqnarray}
The dependence of $\tilde M_{xy}$ on both $R$ and
$R_{\alpha\beta}\Delta^\alpha \Delta^\beta$ in (\ref{e22}) agrees
with that of \cite{r20,r2,r19} (once the different normalization of
$t$ is taken into account).

A completely analogous calculation can be used to fix the lowest
order dependence on $R_{\alpha\beta;\gamma} \Delta^\alpha
\Delta^\beta \Delta^\gamma$; it is found to vanish.  This also
appears to be consistent with the results of \cite{r2} (where
the coefficient
$R_{\alpha\beta;\gamma}$ is evaluated at the end-point $y$, instead
of the mid-point $x_0$).
Further terms in the DeWitt expansion of $\tilde M_{xy}$ can
be similarly determined.

The techniques used in this section may be employed to find the
effective action for a particle moving in a gravitational field
\cite{r28}.  This involves taking $\phi^\alpha(\tau)$ to be
arbitrary in~(\ref{e16}) rather than restricting it to be $x_0$.

\section{Discussion}

In the preceding sections we have considered how the QMPI can be
used to
determine the elements of the DeWitt expansion for the heat kernel
both on and off the diagonal.  This technique is seen to be easier
to use than the original approach \cite{r6} in which the heat
equation was
solved perturbatively.  By employing the off-diagonal elements,
calculations can be done to two-loop order \cite{r9}.  The method
works in both flat and curved space.

Because of the simplicity of the proper-time ($\tau$) integrands
which arise when implementing this method, we expect that
high-order covariant expansions could be computerized using
presently available symbolic algebra
packages.  This approach has already been employed in the
diagonal case~\cite{r7}.  In the
off-diagonal case, surface terms (proportional to $\Delta$)
introduce an additional combinatorical consideration which
should be tractable.

Finally, it is worth noting, that although our treatment of this
problem focused specifically on the case where the vector
potential $A_\mu$ was a gauge field, this is unnecessarily
restrictive and the methods presented here would work equally
well for a general vector coupling to a background field.  For
example, if one were interested in radiative corrections to
Fermionic Green's functions in flat-space QED, the following
quantum mechanical operator would be of interest
\cite{r30}
\begin{equation}
\label{M}
\left(
\begin{array}{ccc}
p^2 g_{\mu\nu} - \left( 1-\case{1}{a} \right) p_\mu p_\nu &
-(\gamma_\mu \psi)^T & (\overline\psi \gamma_\mu) \\
(\gamma_\nu \psi) & 0 & p\!\!\!/ - m \\
-(\overline\psi \gamma_\nu)^T & -(p\!\!\!/ - m)^T & 0
\end{array} \right)_{(\mu,\nu)} \, .
\end{equation}
By factoring in the constant supermatrix
$$
\left( \begin{array}{ccc} g_{\nu\lambda} & 0 & 0 \\
0 & 0 & (p\!\!\!/ + m)^T \\
0 & -(p\!\!\!/ + m) & 0 \end{array}\right) _{(\nu,\lambda)} \, ,
$$
and choosing the $a=1$ gauge, operator (\ref{M}) becomes
\begin{equation}
\left( \begin{array}{ccc} p^2g_{\mu\lambda}
& -(\overline\psi \gamma_\mu)(p\!\!\!/ + m)
& -(\gamma_\mu \psi)^T (p\!\!\!/ + m)^T\\
(\gamma_\lambda \psi)
& p^2+m^2 & 0 \\
-(\overline\psi \gamma_\lambda)^T & 0 & p^2+m^2 \end{array}
\right)_{(\mu,\lambda)} \, .
\end{equation}
After completing the square of $p$ (and noting that $p^T = -p$ in
the coordinate-space representation), the heat kernel of this
operator is easily shown to have the form of equation~(\ref{e1a})
with vector potential,
\begin{equation}
A_\nu(q) = \frac{1}{2}
\left( \begin{array}{ccc} 0 & \overline\psi(q) \gamma_\mu\gamma_\nu
& -(\gamma_\nu \gamma_\mu \psi(q))^T \\
0 & 0 & 0 \\
0 & 0 & 0 \end{array}
\right)_{(\mu,\lambda)} \, ,
\end{equation}
and scalar potential,
\begin{equation}
V(q) = \left( \begin{array}{ccc} 0
& (\overline\psi(q) \gamma_\mu) \left( \case{i}{2}
\loarrow{\partial\!\!\!/} -m \right)
& \left[ \left( \case{-i}{2} \partial\!\!\!/ -m \right)\gamma_\mu
\psi(q) \right]^T \\
(\gamma_\lambda \psi(q))
& m^2 & 0 \\
-(\overline\psi(q) \gamma_\lambda)^T & 0 & m^2
\end{array}
\right)_{(\mu,\lambda)}\, .
\end{equation}
After the appropriate Taylor expansions of these potentials is
substituted into eq.~(\ref{e3}) the method should proceed in the
obvious way.

\section{Acknowledgments}

We would like to thank the Natural Science and Engineering Research
Council of Canada (NSERC) for financial support.

D.G.C.M. would also like to thank the Dublin Institute for Advanced
Study for its hospitality while much of this work was being done.
He would also like to thank C. Wiesendanger for having pointed out
ref.~\cite{r2}.

F.A.D. would like to thank D.J. O'Connor for providing materials
related to appendix~\ref{gauge} and M. L\"uscher for a communication
regarding the results of ref. \cite{r2}.

\appendix
\section{Evaluation of the Path Integral}
\label{PI}

Here we explicitly evaluate the path integrals used
in the previous sections.
Firstly, we define the type of integrals which need to be evaluated
to obtain the expectation value of a function $F$ of dynamical
variables $\xi_\mu(\tau)$ and $b^\mu(\tau)$,
\begin{eqnarray}
\langle F(\xi,b) \rangle_\Delta & \equiv & \left\langle
\case{\Delta}{2},0 \right|
F(\xi,b) \left| -\case{\Delta}{2},0 \right\rangle \nonumber \\
& = & \left[ \frac{e^{-\Delta^\mu \Delta^\nu g_{\mu\nu}/2t}}{(2\pi
t)^{D/2}} \right]^{-1}
\int^{\Delta/2}_{-\Delta/2} D\xi^\alpha(\tau)
\int^0_0 Db_\beta(\tau) \, F(\xi,b)
\label{EVpath} \\
& & \times \exp \left\{ - \int^t_0 d\tau \left[
\case{1}{2} g_{\mu\nu} \dot\xi^\mu(\tau) \dot\xi^\nu(\tau) +
\case{1}{2} g^{\mu\nu} b_\mu (\tau) b_\nu(\tau) \right]
\right \} \nonumber \, ,
\end{eqnarray}
which satisfies the normalization condition
$\langle 1 \rangle_\Delta = 1$.

We eliminate the boundary parameter $\Delta$ from the path integral
by integrating over fluctuations about the classical geodesic
$\xi_{\text{cl}}(\tau) = (-1/2 + \tau/t) \Delta$.  Letting $\xi
\rightarrow \xi + \xi_{\text{cl}}$ it is easily shown that
\begin{equation}
\label{Delta0}
\langle F[\xi,b] \rangle_\Delta =
\langle F[\xi + \xi_{\text{cl}},b] \rangle_0 \, .
\end{equation}
Thus, for the purpose of evaluating any particular term in equation
(\ref{e18}), it will be sufficient for us to concentrate on
evaluating $\langle F[\xi,b] \rangle_0$ where $F$ is a monomial
in $\xi$,
$\dot\xi$ and $b$.  We can follow the procedure usually used in
applying the path integral in field theory and so we only need to
evaluate the
various two-point functions, $\langle b_\mu (\tau) b_\nu (\tau')
\rangle$, $\langle \xi^\mu (\tau) \xi^\nu (\tau') \rangle$, etc,
and then apply the appropriate Wick expansions.

Jumping directly into the continuum limit, one would obtain, using
the standard techniques \cite{r12,r23},
\begin{mathletters}
\begin{eqnarray}
\langle \xi^\mu (\tau) \xi^\nu (\tau') \rangle & = & -G(\tau,\tau')
g^{\mu\nu} \\
& = & - \left[ \case{1}{2} |\tau - \tau'| - \case{1}{2} (\tau +
\tau') + \frac{\tau\tau'}{t} \right] g^{\mu\nu} \, ,\nonumber
\end{eqnarray}
\begin{eqnarray}
\langle b_\mu (\tau) b_\nu (\tau') \rangle & = & G^{\text{ghost}}
(\tau,\tau') g_{\mu\nu} \label{naiveG}\\
& = & \delta(\tau-\tau') g_{\mu\nu}\, ,\nonumber
\end{eqnarray}
\begin{eqnarray}
\langle \xi^\mu (\tau) b_\nu (\tau') \rangle & = & 0
\end{eqnarray}
\end{mathletters}%
Upon close inspection of the above propagators one finds that the
method used for finding the explicit form of the Green's functions
is problematic, especially at the end points.  For example, the
defining equation for $G$
$$
\frac{\partial^2}{\partial \tau^2} G(\tau,\tau') =
\delta(\tau-\tau') \, ,
$$
and the corresponding boundary conditions
$$ G(\tau,0) = G(\tau,t) = 0 \, ,$$
are contradictory at the $\tau' =0$ and $\tau' = t$ boundaries.
Furthermore, the explicit form of the ghost propagator of equation
(\ref{naiveG}) does not satisfy its homogeneous boundary condition
at the corners $(0,0)$ and $(t,t)$.

In order to circumvent these difficulties, we adopt the approach of
\cite{r19} where the path integral is taken to be
the limiting case of integration over a finite set of discrete
Fourier modes.  The Fourier expansions
\begin{equation}
\label{Fourier}
\xi^\mu (\tau) = \sum^M_{n=1} \xi^\mu_n \sin \frac{n\pi\tau}{t}
\, , \hspace{1cm}
b_\mu(\tau) = \sum^M_{n=1} b_{\mu n} \sin \frac{n\pi\tau}{t} \, ,
\end{equation}
automatically incorporate the necessary boundary conditions in the
$\Delta = 0$ path integral; furthermore, they ensure that the path
integral is over functions which are periodic, as required in
\cite{r10}.  If the cutoff $M$ is finite then the
path integral over Fourier coefficients is well defined.
Substituting these expansions into the action of (\ref{EVpath}), we
find that the propagators of the Fourier modes are simply \cite{r19}
\begin{equation}
\langle \xi^\mu_n \xi^\nu_m \rangle = \frac{2t}{n^2 \pi^2}
\delta_{nm} g^{\mu\nu} \, , \hspace{1cm}
\langle b_{\mu n} b_{\nu m} \rangle = \frac{2}{t}
\delta_{nm} g_{\mu\nu} \, ,
\end{equation}
which we take to be our fundamental propagators.  Then using
(\ref{Fourier}) the field propagators are found to be
\begin{mathletters}
\label{FieldProp}
\begin{eqnarray}
\langle \xi^\mu(\tau) \xi^\nu(\tau') \rangle & = & g^{\mu\nu} t
\sum^M_{n=1} \frac{1}{n^2 \pi^2} \left[ \cos \frac{n\pi
(\tau-\tau')}{t} - \cos \frac{n\pi (\tau + \tau')}{t} \right] \, ,
\label{xixi} \\
\langle b_\mu(\tau) b_\nu(\tau') \rangle & = & g_{\mu\nu} \frac{1}{t}
\sum^M_{n=1} \left[ \cos \frac{n\pi
(\tau-\tau')}{t} - \cos \frac{n\pi (\tau + \tau')}{t} \right] \, ,\\
\langle \xi^\mu(\tau) b_\nu(\tau') \rangle & = & 0 \, .
\end{eqnarray}
\end{mathletters}%
These are the propagators which should be substituted into any Wick
expansion of $\langle F[\xi,b] \rangle_0$.  In principal we should
attempt to approximate the continuum limit by letting $M \rightarrow
\infty$ only after the proper-time integrals have been performed.
Unfortunately, these harmonic expressions for the Greens functions
can be unnecessarily cumbersome; in practice whenever the
appropriate
product of Green's functions is sufficiently well behaved as
$M \rightarrow \infty$, we will sum them in that limit before
substituting them into the Wick expansion.
This is most easily done by using the variables
$\tau_\pm = \tau\pm \tau'$.  Then, on the relevant intervals
$\tau_+ \in [0,2t]$, \mbox{$\tau_- \in [-t,t]$}, the following
limits are easily derived from elementary Fourier theory:
\begin{mathletters}
\begin{eqnarray}
\sum^M_{n=1} \frac{t}{n^2\pi^2} \cos \frac{n\pi\tau_\pm}{t}
&\stackrel{M \rightarrow \infty}{\longrightarrow}&
\case{1}{6} t -
\case{1}{2} | \tau_\pm | + \frac{1}{4t} \tau_\pm^2
\\
\label{sin-}
\sum^M_{n=1} \frac{1}{n\pi} \sin \frac{n\pi \tau_-}{t}
&\stackrel{M \rightarrow \infty}{\longrightarrow}&
\case{1}{2} {\rm sgn}{\tau_-} - \frac{1}{2t} \tau_-
\\
\sum^M_{n=1} \frac{1}{n\pi} \sin \frac{n\pi \tau_+}{t}
&\stackrel{M \rightarrow \infty}{\longrightarrow}&
\case{1}{2} - \frac{1}{2t} \tau_+
\\
\label{cos-}
\sum^M_{n=1} \frac{1}{t} \cos \frac{n\pi\tau_-}{t}
&\stackrel{M \rightarrow \infty}{\longrightarrow}&
\delta(\tau_-) - \frac{1}{2t}
\\
\label{cos+}
\sum^M_{n=1} \frac{1}{t} \cos \frac{n\pi\tau_+}{t}
&\stackrel{M \rightarrow \infty}{\longrightarrow}&
\delta(\tau_+) +
\delta(2t-\tau_+) - \frac{1}{2t}
\end{eqnarray}
\end{mathletters}%
It is thus found that in many cases the following
prescription is sufficient to recover those propogators which
arise in (\ref{e18}):
\begin{mathletters}
\label{summedProp}
\begin{eqnarray}
\langle \xi^\mu(\tau) \xi^\nu(\tau') \rangle & \rightarrow &
-G(\tau,\tau')
\equiv - \left[ \case{1}{2}|\tau-\tau'| - \case{1}{2}
(\tau + \tau') +
\case{\tau\tau'}{t} \right] g^{\mu\nu} \label{G} \\
\langle \dot\xi^\mu(\tau) \xi^\nu(\tau') \rangle & \rightarrow &
-G(\dot\tau,\tau')
\equiv - \left[ \case{1}{2} {\rm sgn}(\tau-\tau') - \case{1}{2} +
\case{\tau'}{t} \right] g^{\mu\nu} \label{Gdot}\\
\langle \dot\xi^\mu(\tau) \dot\xi^\nu(\tau') \rangle &
\rightarrow &
-G(\dot\tau,\dot\tau')
\equiv - \left[ -\delta(\tau-\tau') +  \case{1}{t} - \delta(2t
-\tau-\tau') - \delta(\tau + \tau') \right] g^{\mu\nu}
\label{Gdotdot} \\
\langle b_\mu(\tau) b_\nu(\tau') \rangle & \rightarrow &
G^{\text{ghost}}(\tau,\tau')
\equiv \left[ \delta(\tau-\tau') - \delta(2t-\tau-\tau') -
\delta(\tau + \tau') \right] g_{\mu\nu} \label{Gghost}
\end{eqnarray}
\end{mathletters}%
(We note here that (\ref{Gdotdot}) can be derived from (\ref{G}) by
requiring that $G$ be extended periodically outside of the domain $0
\leq \tau, \tau' \leq t$,
i.e $G(\tau+t,\tau'+t) \equiv G(\tau,\tau')$.  This periodicity is
akin to the definition of the Green's function given in \cite{r10}
on a compact periodic
surface.  A discussion of this point can be found in
\cite{r14}.)

Of note is the limiting case where $\tau = \tau'$
which appears in the calculation of (\ref{e20}); there the
following considerations apply:
\begin{itemize}
\item{The term ${\rm sgn}(\tau-\tau')$ in (\ref{Gdot}) is odd in
$\tau-\tau'$ so its Fourier transform necessarily vanishes in the
$\tau=\tau'$ limit.
Thus we should take ${\rm sgn}(0) = 0$ (c.f. (\ref{sin-})).}
\item{In the $\tau=\tau'$ limit, the Fourier series of
$\delta(\tau-\tau')$ in (\ref{Gdotdot}) and (\ref{Gghost}) leads
to a regulated representation of $\delta(0)$ given by
$t\delta(0) = M + 1/2$, (c.f. (\ref{cos-})).}
\item{Since the Fourier series of the terms $\delta(2t-2\tau) +
\delta(2\tau)$ (which arise in (\ref{Gdotdot}) and (\ref{Gghost}))
is even about the respective poles
(c.f. (\ref{cos+})), any integrated contribution
from these terms should be halved as a result of the poles
coinciding with the end-point of the integration region, i.e.
$$
\int^t_0 d\tau \left[ \delta(2t-2\tau) + \delta(2\tau)
\right] f(\tau) = \case{1}{4} \left[ f(t) + f(0) \right] \, .
$$}
\end{itemize}

\section{Normal Coordinate Expansion of the Gauge Field }
\label{gauge}

In this appendix we discuss the construction of a gauge-covariant
normal coordinate expansion for the gauge potential.

By analogy with the flat-space case discussed briefly in
section~\ref{flat}
and in refs.~\cite{r11,r7}, the appropriate gauge condition for this
expansion is the {\it synchronous} gauge~\cite{r29} (a curved-space
generalization of the Fock-Schwinger gauge~(\ref{e5})~\cite{r22})
which fits very well
in the normal coordinate construction.  In the basis of the normal
coordinate system, the gauge condition is
\begin{equation}
\xi^\alpha A_\alpha(\phi + \pi(\xi)) = 0 \, .
\end{equation}
Either by integrating along the
geodesics (which is formally identical to equation~(\ref{e6a})) or
by using differential forms~\cite{r29}
one can show, in the normal coordinate system, that the synchronous
gauge leads to a gauge-covariant expansion for the vector potential
which looks exactly like equation~(\ref{e6b}) with the
gauge-covariant normal coordinate derivative
$D_\alpha = \frac{\partial}{\partial \xi^\alpha} + [A_\alpha,
\ldots]$.
The latter derivative is not covariant under reparametrization of
the manifold, however using the methods of
reference~\cite{r26} it is straightforward to write such
normal coordinate derivatives at the origin in terms of the
corresponding fully-covariant derivatives, denoted by indices
trailing the semicolon ($;$).  For example, one can show that
\footnote{The authors suspect that the fourth derivative of
a rank-two tensor implied in reference~\cite{r26} is not
entirely correct.  The corresponding coefficients presented here
for the field strength, equation~(\ref{DDDDF}),
have been verified independently.}
\begin{mathletters}
\label{DDDDDF}
\begin{eqnarray}
D_{\beta_1} F_{{\beta_0}\gamma} & \dot = &
F_{{\beta_0}\gamma;\beta_1}
\label{DF} \\
D_{\beta_2} D_{\beta_1} F_{{\beta_0}\gamma} & \dot = &
F_{{\beta_0}\gamma;\beta_1\beta_2}
+ \case{1}{3} F_{{\beta_0}\delta}
R^\delta_{\beta_1 \beta_2 \gamma} \label{DDF} \\
D_{\beta_3} D_{\beta_2} D_{\beta_1} F_{{\beta_0}\gamma}
& \dot = &
F_{{\beta_0}\gamma;\beta_1\beta_2\beta_3}
+ \case{1}{2}
F_{{\beta_0}\delta}R^\delta_{\beta_1\beta_2\gamma;\beta_3}
+ F_{{\beta_0}\delta;\beta_1}
R^\delta_{\beta_2 \beta_3 \gamma}
\label{DDDF} \\
D_{\beta_4} D_{\beta_3} D_{\beta_2} D_{\beta_1}
F_{{\beta_0}\gamma} & \dot = &
F_{{\beta_0}\gamma;\beta_1\beta_2\beta_3\beta_4}
+ \case{3}{5} F_{{\beta_0}\delta}
R^\delta_{\beta_1\beta_2\gamma;\beta_3\beta_4}
+ 2 F_{{\beta_0}\delta;\beta_1}
R^\delta_{\beta_2\beta_3\gamma;\beta_4}
\nonumber \\
& & \hspace{1cm}
+ 2 F_{{\beta_0}\delta;\beta_1\beta_2}
R^\delta_{\beta_3\beta_4\gamma}
+ \case{1}{5} F_{{\beta_0}\epsilon}
R^\epsilon_{\beta_1\beta_2\delta} R^\delta_{\beta_3\beta_4\gamma}
\label{DDDDF}
\end{eqnarray}
\end{mathletters}%
where $\dot =$ indicates equality at the origin
only after symmetrization of the $\beta_i$ indices.  Substitution
of equations~(\ref{DDDDDF}) into eq.~(\ref{e6b})
yields the fully covariant normal coordinate expansion to
fifth-order in the normal coordinates,
\begin{eqnarray}
\label{AexpansionF}
A_\gamma(\phi + \pi(\xi)) & = &
\case{1}{2}\left\{ F_{\beta\gamma} \right\} \xi^\beta
+ \case{1}{3} \left\{ F_{{\beta_0}\gamma;\beta_1}
\right\} \xi^{\beta_0} \xi^{\beta_1}
+ \case{1}{8} \left\{F_{{\beta_0}\gamma;\beta_1\beta_2}
+ \case{1}{3} F_{{\beta_0}\delta} R^\delta_{\beta_1 \beta_2 \gamma}
\right\} \xi^{\beta_0} \xi^{\beta_1}
\xi^{\beta_2} \nonumber \\
& & + \case{1}{3! 5}
\left\{F_{{\beta_0}\gamma;\beta_1\beta_2\beta_3}
+ \case{1}{2} F_{{\beta_0}\delta}
R^\delta_{\beta_1\beta_2\gamma;\beta_3}
+ F_{{\beta_0}\delta;\beta_1}R^\delta_{\beta_2 \beta_3 \gamma}
\right\}
\xi^{\beta_0} \xi^{\beta_1}\xi^{\beta_2}\xi^{\beta_3} \nonumber \\
& & + \case{1}{4! 6}
\left\{F_{{\beta_0}\gamma;\beta_1\beta_2\beta_3\beta_4}
+ \case{3}{5}
F_{{\beta_0}\delta} R^\delta_{\beta_1\beta_2\gamma;\beta_3\beta_4}
+ 2 F_{{\beta_0}\delta;\beta_1}
R^\delta_{\beta_2\beta_3\gamma;\beta_4}
\right. \\
& & \left. \;\;\;\;\;\;\;\;\;\;
+ 2 F_{{\beta_0}\delta;\beta_1\beta_2}
R^\delta_{\beta_3\beta_4\gamma}
+ \case{1}{5} F_{{\beta_0}\epsilon}
R^\epsilon_{\beta_1\beta_2\delta} R^\delta_{\beta_3\beta_4\gamma}
\right\}
\xi^{\beta_0} \xi^{\beta_1} \xi^{\beta_2} \xi^{\beta_3}
\xi^{\beta_4} \nonumber \\
& & + {\cal O} \left(\xi^6\right) \, .\nonumber
\end{eqnarray}
All coefficients in braces $\{ \cdots \}$ are evaluated at the
origin, where the basis vectors for the normal coordinate system
coincide with those of the original system.
Since the potential on the left hand side of this equation is
not a vector at the origin, its indices must refer to the
normal coordinate basis.
(This is also true of equations~\mbox{(\ref{e17b})-(\ref{e17c})}).
The results of (\ref{AexpansionF}) agree with those of~\cite{r2}
to order ${\cal O} \left( \xi^3 \right)$.

%
\mediumtext
\begin{table}
\caption{Possible contributions to the various coefficients of the
DeWitt expansion.}\label{t1}
\begin{tabular}{cl}
Coefficient & Contributions \\ \hline\hline
$a_0\left(x_0,\Delta\right)$ &$1$ \\ \hline
$a_1\left(x_0,\Delta\right)$ &$\left(\Delta \cdot D\right)^k V,
\left(\Delta \cdot D\right)^k \left(\Delta^\alpha D_\beta
{F_\alpha}^\beta\right),
\left(\Delta \cdot D\right)^k
\left(\Delta^\alpha \Delta^\beta F_{\alpha\mu} {F_\beta}^\mu\right)
$ \\ \hline
$a_2\left(x_0,\Delta\right)$ &$
\left(\Delta\cdot D\right)^k \left(D^2 V\right),
\left(\Delta \cdot D\right)^{k}V\left(\Delta \cdot D\right)^{l}V,
$\\
& $ \left(\Delta\cdot D\right)^{k} V \left(\Delta\cdot
D\right)^{l}\left(\Delta^\alpha D_\beta
{F_\alpha}^\beta\right),
\left(\Delta\cdot D\right)^{k}\left(\Delta^\alpha D_\beta
{F_\alpha}^\beta\right)\left(\Delta\cdot
D\right)^{l} V ,
$ \\
& $
\left(\Delta\cdot D\right)^{k}\left(\Delta^\alpha \Delta^\beta
F_{\alpha\mu}
{F_\beta}^\mu\right)\left(\Delta\cdot D\right)^{l} V,
\left(\Delta\cdot D\right)^{k} V
\left(\Delta\cdot D\right)^{l}\left(\Delta^\alpha \Delta^\beta
F_{\alpha\mu}{F_\beta}^\mu\right), $ \\
& $ \left(\Delta \cdot D\right)^{k}F^{\alpha\beta}\left(\Delta
\cdot D\right)^{l}F_{\alpha\beta},$\\
& $ \left(\Delta\cdot D\right)^{k} \left(\Delta^\alpha
F_{\alpha\mu} \right) \left(\Delta\cdot
D\right)^{l}\left(D_\beta F^{\beta\mu}\right) ,
 \left(\Delta\cdot D\right)^{k}\left(D_\beta F^{\beta\mu}\right)
\left(\Delta\cdot D\right)^{l}
\left(\Delta^\alpha F_{\alpha\mu} \right), $ \\
& $ \left(\Delta \cdot D\right)^{k} \left(\Delta^\alpha D_\beta
{F_\alpha}^\beta\right)
\left(\Delta \cdot D\right)^{l} \left(\Delta^\alpha D_\beta
{F_\alpha}^\beta\right),$\\
& $
\left(\Delta\cdot D\right)^{k} \left(\Delta^\alpha D_\beta
{F_\alpha}^\beta\right)\left(\Delta\cdot
D\right)^{l}\left(\Delta^\alpha \Delta^\beta
F_{\alpha\mu}{F_\beta}^\mu\right), $ \\
& $
\left(\Delta\cdot D\right)^{k}\left(\Delta^\alpha \Delta^\beta
F_{\alpha\mu}{F_\beta}^\mu\right)\left(\Delta\cdot D\right)^{l}
\left(\Delta^\alpha D_\beta
{F_\alpha}^\beta\right), $ \\
& $
\left(\Delta \cdot D\right)^{k} \left( \Delta^\alpha
\Delta^\beta F_{\alpha\mu} {F_\beta}^\mu \right)
\left(\Delta \cdot D\right)^{l} \left( \Delta^\alpha
\Delta^\beta F_{\alpha\mu} {F_\beta}^\mu \right)
$ \\ \hline
\vdots & etc. \\
\end{tabular}
\centering
(All fields and covariant derivatives are evaluated
at $x_0$.)\\
$ (k,l = 0,1,2, \cdots) $
\end{table}

\begin{thebibliography}{99}
\bibitem{r1} B. DeWitt, {\it Dynamical Theory of Groups and Fields},
(Gordon and Breach, New York 1965). \\
H.P. McKean and J.M. Singer, J. Diff. Geom. {\bf 5}, 233 (1971). \\
R.T. Seeley, Amer. Math. Soc. {\bf 10}, 288 (1967).\\
P.B. Gilkey, J. Diff. Geom. {\bf 10}, 601 (1975).
\bibitem{r2} M. L\"uscher, Ann. Phys. {\bf 142}, 359 (1982).
\bibitem{r3} V.P. Gusynin, Nucl. Phys. B{\bf 333}, 296 (1990).
\bibitem{r4} R. Nepomechie, Phys. Rev. D{\bf 31}, 3291 (1985). \\
M. Reuter, Phys. Rev. D{\bf31}, 1374 (1985).
\bibitem{r5} R.B. Mann, D.G.C. McKeon, T. Steele and L. Tarasov,
Nucl. Phys. B{\bf 311}, 630 (1989).
\bibitem{r6} L. Culumovic and D.G.C. McKeon, Phys. Rev. D{\bf38},
3831 (1988).
\bibitem{r7} D. Fliegner, P. Haberl, M.G. Schmidt and C. Schubert,
DESY preprint 94-221, \mbox{hep-th/9411177}. \\
{\it ibid}, \mbox{hep-th/9505077}.
\bibitem{r8} J. Beckenstein and L. Parker, Phys. Rev. D{\bf 19},
483 (1979).
\bibitem{r9} L. Culumovic, D.G.C. McKeon and T.N. Sherry, Ann. of
Phys. {\bf 197}, 94 (1989). \\
L. Culumovic and D.G.C. McKeon, Can. J. Phys. {\bf 68}, 1166
(1990).\\
D.G.C. McKeon and S.K. Wong, Int. J. Mod. Phys. A{\bf 10}, 2181
(1995).\\
F.A. Dilkes and D.G.C. McKeon, Phys. Rev. D{\bf 52}, 4668 (1995),
\mbox{hep-th/9502075}.
\bibitem{r10} M.J. Strassler, Nucl. Phys. B{\bf 385}, 145 (1992).\\
A.M. Polyakov, {\it Gauge Fields and Strings}
 section 9.3 (Harwood Academic, Chur 1987).
\bibitem{r11} M.G. Schmidt and C. Schubert, Phys. Lett. B{\bf318},
438 (1993), hep-th/9309055. \\
{\it ibid}, hep-ph/9412358.\\
D. Flieger, M.G. Schmidt and C. Schubert, Z. Phys C{\bf 64}, 111
(1994), \mbox{hep-ph/9401221}.
\bibitem{r12} D.G.C. McKeon, Can. J. Phys. {\bf 70}, 652 (1992). \\
D.G.C. McKeon and A. Rebhan, Phys. Rev. D{\bf 48}, 2891 (1993).
\bibitem{r13} D.G.C. McKeon and T.N. Sherry, Mod. Phys. Lett. (in
press), Int. J. of Theor. Phys. {\bf 32}, 1105 (1993).
\bibitem{r16} D.G.C. McKeon and S.K. Wong, J. Math. Phys. {\bf 36},
1691 (1995).
\bibitem{r14} D.G.C. McKeon, Ann. of Phys. {\bf 224}, 139 (1993).
\bibitem{r15} M.G. Schmidt and C. Schubert, Phys. Lett. B{\bf 331},
69 (1994).
\bibitem{r17} B.S. DeWitt, Rev. Mod. Phys. {\bf 29}, 377 (1957). \\
K.S. Chen, J. Math. Phys. {\bf 13}, 1723 (1972). \\
J. Dowker, J. Phys. A{\bf 7}, 1256 (1974). \\
G.A. Ringwood, J. Phys. A{\bf 9}, 1253 (1976).
\bibitem{r18} M.S. Marinov, Phys. Rep. {\bf 60}, 1 (1980). \\
D.C. Khandakar, S.V. Lawande and K.V. Bhagwat, {\it Path Integral
Methods and their Applications} (World Scientific, Singapore 1993).
\bibitem{r19} F. Bastianelli, Nucl. Phys. B{\bf 376}, 113 (1992),
\mbox{hep-th/9112035}. \\
F. Bastianelli and P. Van Nieuwenhuizen, Nucl. Phys. B{\bf 389}, 53
(1993), \mbox{hep-th/9208059}.
\bibitem{r20} L. Parker and D. Toms, Phys. Rev. D{\bf 31}, 953
(1985). \\
I. Jack and L. Parker, Phys. Rev. D{\bf 31}, 2439 (1985).
\bibitem{r21} R.P. Feynman, Rev. Mod. Phys. {\bf 20}, 367 (1948).
\bibitem{r22} A. Fock, Phys. Z. Sowjetunion {\bf 12}, 404 (1937). \\
C. Cr\"onstrom, Phys. Lett. {\bf 90B}, 267 (1980). \\
M.A. Schifman, Nucl. Phys. B{\bf 173}, 13 (1980).
\bibitem{Igor} V.P. Gusynin and I.A. Shovkovy, Preprint UG - 9/95,
\mbox{hep-ph/9509383}.
\bibitem{r23} C. Itzykson and J.B. Zuber, {\it Quantum Field
Theory} (McGraw-Hill, New York, 1980).
\bibitem{r24} L. Parker, in {\it Recent Developments in
Gravitation, Carg\`{e}se 1978} (ed. M. Levy and S. Deser, Plenum
Press, London, 1978).
\bibitem{r25} J. Honerkamp, Nucl. Phys. B{\bf 36}, 130 (1972).
\bibitem{r26} L. Alvarez-Gaum\'{e}, D.Z. Freedman and S. Mukhi,
Ann. of Phys. {\bf 134}, 85 (1981).
\bibitem{r32} M. Leblanc, R.B. Mann, D.G.C. McKeon and T.N. Sherry,
Nucl. Phys. B{\bf 349}, 494 (1991).
\bibitem{r27} O. Veblen, {\it Invariants of Quadratic Differential
Forms} (Cambridge Press, Cambridge 1927). \\
L. Eisenhart, {\it Riemann Geometry} (Princeton 1965). \\
A.Z. Petrov, {\it Einstein Spaces} (Pegamon Press, New York, 1969).
\bibitem{r28} D.G.C. McKeon, Can. J. Phys. {\bf 67}, 837 (1989).
\bibitem{r30} D.G.C. McKeon and T.N. Sherry, Phys. Rev. D{\bf 35},
3854 (1987).\\
D.G.C. McKeon, Mod. Phys. Lett. A{\bf 6}, 3711 (1991).
\bibitem{r29} D.J. O'Connor, PhD dissertation (University of
Maryland), 1985. \\
D.J. O'Connor, Nucl. Phys. B{\bf 298}, 429 (1988).
\end{thebibliography}
\end{document}